\documentclass[12pt]{iopart}

\usepackage{iopams}
\usepackage{graphicx}
\usepackage{caption2}
\usepackage{subfigure}
\usepackage{float}
\usepackage{braket}
\usepackage{stfloats}
  
\usepackage{color}
\usepackage{amsfonts}
\usepackage{autobreak}
\usepackage{multirow}
\usepackage{makecell}
\usepackage{cite}

\begin{document}

\title
{Novel Architecture of Parameterized Quantum Circuit for  Graph Convolutional Network}

\author{Yanhu Chen$^1$, Cen Wang $^2$, Hongxiang Guo$^{1,*}$, Jianwu $^1$}

\address{$^1$ \emph{Beijing University of Posts and Telecommunications, Beijing, China,}}
\address{$^2$ KDDI \emph{Research, Inc.}}
\ead{yanhuchen@bupt.edu.cn;xce-wang@kddi.com; hxguo@bupt.edu.cn;jianwu@bupt.edu.cn}
\vspace{10pt}
\begin{indented}
\item[]Feb 2022
\end{indented}

\begin{abstract}
Recently, the implementation of quantum neural networks is based on noisy intermediate-scale quantum (NISQ) devices. Parameterized quantum circuit (PQC) is such the method, and its current design just can handle linear data classification. However, data in the real world often shows a topological structure. In the machine learning field, the classical graph convolutional layer (GCL)-based graph convolutional network (GCN) can well handle the topological data. Inspired by the architecture of a classical GCN, in this paper, to expand the function of the PQC, we design a novel PQC architecture to realize a quantum GCN (QGCN). More specifically, we first implement an adjacent matrix based on linear combination unitary and a weight matrix in a quantum GCL, and then by stacking multiple GCLs, we obtain the QGCN. In addition, we first achieve gradients decent on quantum circuit following the parameter-shift rule for a GCL and then for the QGCN. We evaluate the performance of the QGCN by conducting a node classification task on Cora dataset with topological data. The numerical simulation result shows that QGCN has the same performance as its classical counterpart, the GCN, in contrast, requires less tunable parameters. Compared to a traditional PQC, we also verify that deploying an extra adjacent matrix can significantly improve the classification performance for quantum topological data.

\end{abstract}

%
%
%
%
%

\section{Introduction}

Quantum neural networks (QNN) adopt quantum frameworks to realize the functions of classical deep learning (DL) models. QNN has shown several advantages including reducing the complexity of machine learning models, expanding the data scale, and reducing the number of trainable parameters \textcolor{red}{\cite{levine2019quantum, carrasquilla2017machine}}. Implementing QNN mainly has two approaches: (1) based on a universal fault-tolerant quantum computer, a quantum algorithm is used to accelerate a key calculation step of a classical DL model (e.g., 
vector multiplication and convolution) \textcolor{red}{\cite{chenaccelerating, wei2022quantum, chen2020quantum,  chen2021hybrid}}; (2) based on a current  noisy intermediate-scale quantum (NISQ) device \textcolor{red}{\cite{mitarai2018quantum, benedetti2019parameterized}}, an entire classical DL model is replaced with a quantum structure suitable for the device, although the function is the same, the implementation logic is different.

Parameterized quantum circuit (PQC) is such a mainstream quantum structure that is easy to deploy on NISQ-like devices, and currently achieves the same function with a fully connected neural network. A typical PQC is composed of multiple fixed quantum gates (e.g., the Pauli-Z gate and the controlled NOT gate) and multiple adjustable quantum gates (e.g., (controlled) $R_y(\theta)$ gate), and the input data usually is encoded into quantum amplitudes \textcolor{red}{\cite{hur2021quantum, araujo2021divide}}. In a classical DL, each element of a weight matrix is the trainable parameter, whereas an unitary matrix corresponding to PQC is equivalent to a weight matrix. The unitary matrix is not directly given and is obtained by matrix multiplications or Kronecker products between the fixed and adjustable quantum gates. Due to such a calculation process, although the source of the element change of the unitary matrix is the phase $\theta$ in an adjustable quantum gate, the effect of the phase change of the quantum gate on the unitary matrix is indirect. Therefore, the type and arrangement of the fixed and the adjustable quantum gates determine the structure of a PQC, as they determine the generation of the unitary matrix. Several evaluations have verified significant impact of a PQC architecture on its performance~\textcolor{red}{\cite{grimsley2019adaptive, benedetti2019generative, benedetti2019adversarial, linke2017experimental, zoufal2019quantum, chivilikhin2020mog, zhang2020differentiable, du2020quantum}}. By an optimization method (e.g., gradient descent algorithm, Nelder-mead, and the heuristic algorithm \textcolor{red}{\cite{psr, nelder-mead, NSGA-II}}), a PQC learns the best unitary matrix for a task.

Although current PQC ensures the shallow circuit depth to fit the NISQ devices, to further expand its functions, PQC should also learn from classical deep learning algorithms, especially for the learning ability of different data types. Some data in the real world are inherently topological. For example, in e-commerce, interactions between users and commodities are topological data that supports highly accurate recommendations. In chemistry, molecules are modeled as graphs, and their bioactivity is then identified for drug discovery. In a citation network, papers are linked to each other via citationships which is an adjacent matrix and helps to achieve paper categorization \textcolor{red}{\cite{wu2020comprehensive}}. To handle such topological data, the current PQC architecture needs a redesign. In addition, in the classical DL algorithm, since the graph convolutional layer shows good performance to adapt topological data, our design goal can be to use PQC to achieve the same function as GCL.

Considering these, in this paper, by redesigning the PQC, we propose quantum graph convolutional layer (QGCL). More specifically, (1) we first design a quantum circuit for an adjacent matrix which is introduced in a classical GCL in addition to a weight matrix~\textcolor{red}{\cite{GCN}}. This adjacent matrix is the key to extract and aggregate the features from each node and its neighboring nodes; (2) then, we use a series of adjustable quantum gates with tunable phases to help realize a proper weight matrix. The challenge is in the first step, because an adjacent matrix is usually not unitary. One cannot directly implement an adjacent matrix on a quantum circuit. Our solution is to use to linear combination unitary (LCU) operator~\textcolor{red}{\cite{gui2006general, childs2012hamiltonian}}, where multiple non-unitary operators are embedded in the higher-dimensional Hilbert space to realize a locally non-unitary yet holistically unitary operator. Therefore, we decompose an adjacency matrix into the sum of several unitary matrices, and realize the adjacency matrix indirectly by implementing these unitary matrices on a quantum circuit. By stacking QGCL, we successfully construct QGCN. Moreover, to optimize parameters, we propose a training scheme based on the parameter-shift rule (PSR) \textcolor{red}{\cite{psr}} to achieve the gradient descent. The advantage to adopt PSR is that we can use the QGCN quantum circuit to realize gradient decent by only changing the phases. In this way, QGCN does not slow down the entire model. The slowdown, in other quantum models, is mainly caused by to the use classical computers for gradients decent. When the amount of data increases, such the architectural advantage of QGCN will be more obvious.

Our contributions are summarized as follows:  
\begin{enumerate} 
\item [(1)] Based on LCU operator, heuristic algorithm, and the design method of reversible logic circuit, we successfully realize the adjacent matrix (non-unitary matrix) on the quantum circuit.  

\item [(2)] Based on PSR, we achieve to update the tunable parameters in QGCN on the quantum circuit rather than on the classical computer, which guarantee the model speed.

\item [(3)] By newly designing the PQC architecture where quantum circuits for adjacent matrix and weight matrix are integrated, we are the first to propose QGCL, and then construct QGCN using multiple QGCLs.
\end{enumerate}

We apply our QGCN on a real citation network dataset, Cora . The numerical simulation results show that QGCN has the same performance as classical GCN but requires less number of parameters. Our work also suggests that a novel PQC architecture for QGCN extends the function of the representative PQC, that to effectively classify quantum data with topological structure.

\section{Problem Statement}
Given an undirected graph $G=(V,E)$, where $V$ denotes the set of nodes, and $E$ denotes the set of edges. $G$ has a total of $N$ nodes. Each node has a label for training and a feature with $D$ dimensions. Each edge does not have a feature. We divide the node set into training set and test set, and we delete the labels for the test nodes. Our task is the node classification, and the optimization goal is to correctly predict a label for a test node.

\section{Method}
\subsection{Graph Convolutional Layer}
\label{sec_GCL}
GCL is a single layer and a key step of GCN. To describe the implementation of GCL by quantum computing, we briefly introduce the classical GCL~\textcolor{red}{\cite{GCN}}. The definition of a GCL is as shown in Equation~\ref{GCL}: 

\begin{equation}
\mathcal{X}^{(l+1)} = \sigma(\hat{A}\mathcal{X}^{(l)}W^{(l)})
\label{GCL}
\end{equation}


\noindent where, $\hat{A}=\widetilde{D}^{-\frac{1}{2}}\widetilde{A}\widetilde{D}^{-\frac{1}{2}}$; $\widetilde{A}=A+I_N$ is the adjacent matrix of the undirected graph $\mathcal{G}$ with self-connections added; $A \in \mathbb{R}^{N\times N}$, and $I_N$ is the identity matrix; $\widetilde{D}$ is a diagnose matrix, and $D_{(i,i)}=\sum_j \widetilde{A}_{i,j}$. $W^{(l)}$ denotes the $l$-th learnable weight matrix, and $W^{(l)}\in R^{D^{(l)} \times D^{(l+1)} }$. $\mathcal{X}^{(l)}$ denotes the input matrix of the $l$-th layer, and $\mathcal{X}^{(l)}\in \mathbb{R}^{N \times D^{(l)} }$; in particular, $\mathcal{X}^{(0)}=\mathcal{X}$; $\mathcal{X}$ denotes the input matrix composed of the input vectors of all nodes: $\mathcal{X} \in \mathbb{R}^{N \times D^{(0)}}$, and $D^{(0)}=D$. $\sigma (\cdot)$ denotes a nonlinear activation function. 



\subsection{The Quantum Circuit for the Adjacent Matrix}
As aforementioned, the adjacent matrix $\hat{A}$ is a real symmetric matrix and usually not a unitary matrix. We use several unitary matrices $\mathcal{U}_k \in \mathbb{R}^{N \times N}$ to approximate the adjacent matrix $\hat{A}$, as shown in Equation \eqref{U_k}:

\begin{equation}
\hat{A}=\sum_k h_k\mathcal{U}_k+ \Delta
\label{U_k}
\end{equation}

\noindent where, $h_k$ represents the weight of the $k$-th unitary matrix $\mathcal{U}_k$ in $\hat{A}$, $\Delta$ is the residual term. Because the Equation \eqref{U_k} requires an arithmetic addition of unitary matrices, we introduce how to implement this arithmetic addition on a quantum circuit based on the linear combination unitary (LCU) operator. We start with the addition of two unitary matrices and generalize to multiple unitary matrices. 

\begin{figure}
	\centering
		\includegraphics[width=8cm]{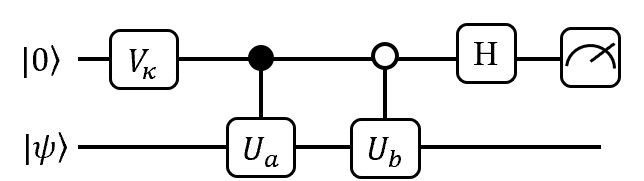}
	\caption{The quantum circuit of LCU to realize $U_a+U_b$.}
	\label{qcLCU}
\end{figure}

Consider a quantum system having an ancillary qubit, working quantum register, and two unitary matrices $U_a$ and $U_b$,  we implement $U_a+U_b$ on a quantum circuit, as shown in \textcolor{red}{Fig \ref{qcLCU}}, where $V_\kappa$ is detailed in Equation~\ref{vk}.

\begin{equation}
V_\kappa=\left[ \begin{matrix}
\sqrt{\frac{\kappa}{\kappa+1}} & \frac{-1}{\sqrt{\kappa+1}}\\
\frac{1}{\sqrt{\kappa+1}} & \sqrt{\frac{\kappa}{\kappa+1}}
\end{matrix}\right]
\label{vk}
\end{equation}

The quantum state evolution process is shown in Equation \eqref{LCU}. When the measurement result of the ancillary qubit is 0, $U_a+U_b$ is successfully applied on $\ket{\psi}$. The successful probability of the quantum circuit as shown in~\textcolor{red}{Fig \ref{qcLCU}} is $1/2$. Recursively, by increasing the number of ancillary qubits, we achieve the summation of multiple unitary matrices.

\begin{equation}
\begin{split}
\ket{0}\ket{\psi} \rightarrow
\left(
\sqrt{\frac{\kappa}{\kappa+1}}\ket{0} + 
\frac{1}{\sqrt{\kappa+1}}\ket{1} 
\right)\ket{\psi} \\
\rightarrow \sqrt{\frac{\kappa}{\kappa+1}}\ket{0}U_a\ket{\psi}+\frac{1}{\sqrt{\kappa+1}}\ket{1} U_b\ket{\psi} \\
\rightarrow \frac{\sqrt{2}}{2} \sqrt{\frac{\kappa}{\kappa+1}}
\left(\ket{0}+\ket{1}\right)U_a \ket{\psi}+
\frac{\sqrt{2}}{2} \frac{1}{\sqrt{\kappa+1}}
\left(\ket{0}-\ket{1}\right)U_b \ket{\psi} \\
=\frac{\sqrt{2}}{2}\ket{0}\left(
\sqrt{\frac{\kappa}{\kappa+1}}U_a+\frac{1}{\sqrt{\kappa+1}}U_b\right)\ket{\psi}+
\frac{\sqrt{2}}{2}\ket{1}\left(
\sqrt{\frac{\kappa}{\kappa+1}}U_a-\frac{1}{\sqrt{\kappa+1}}U_b\right)\ket{\psi}
\end{split}
\label{LCU}
\end{equation}

According to the Equation~\ref{U_k}, we focus on the unitary matrix $\mathcal{U}_k$ and omit the residual term $\Delta$. We give two approaches to construct the quantum circuit of $\mathcal{U}_k$. 

\textbf{Approach 1:} The adjacent matrix $\hat{A}$ is regarded as a Hamiltonian $\mathcal{H}$, since $\hat{A}$ is a real symmetric matrix. An arbitrary Hamiltonian $\mathcal{H}$ can always be decomposed into Kronecker products and a summation of Pauli operators~\textcolor{red}{\cite{peruzzo2014variational}}, as shown in Equation \eqref{Hamiltonian}.

\begin{equation}
\mathcal{H}= h + \sum_{i,\alpha} h_{(\alpha)}^i \alpha^{(i)}+
\sum_{i,j,\alpha,\beta} h_{(\alpha,\beta)}^{i,j}\alpha^{(i)} \otimes \beta^{(j)} + \cdots
\label{Hamiltonian}
\end{equation}

\noindent where $h$, $h_\alpha^{(i)}$, $h_{\alpha,\beta}^{(i,j)}$ are real numbers, $\alpha$, $\beta$ denote the Pauli operators, and $\alpha, \beta \in \{X,Y,Z\}$. $\alpha^{(i)}$ denotes that the $\alpha$ operator is applied on the $i$-th qubit. Note that the Kronecker product of the identity matrix is omitted in Equation \eqref{Hamiltonian}. Each term in Equation~\eqref{Hamiltonian} corresponds to each term in Equation~\eqref{U_k}.

\textbf{Approach 2:} We refer to several existing design methods of reversible circuits~\textcolor{red}{\cite{li2007application, maslov2007techniques, zheng2009novel, saeedi2013synthesis}}. More specifically, let $\ket{y_j}=\mathcal{U}_k\ket{x_j}$, where $\ket{x_j}$ and $\ket{y_j}$ denote the input and output state, respectively. $x_j, y_j \in [0,2^n-1]$ are integer numbers, and $n=\log_2 N$. Because $\mathcal{U}_k$ is a permutation matrix, $x_j$ and $y_j$ denote the element position of the $x_j$-th column and the $y_j$-th row in $\mathcal{U}_k$, respectively, and the element is not equal to 0. We use binary to represent $x_j$ and $y_j$ as shown in Equation~\eqref{x_j} and Equation~\eqref{y_j}, respectively.

\begin{equation}
x_j=x_j^{(n-1)}x_j^{(n-2)} \cdots x_j^{(0)}
\label{x_j}
\end{equation}

\begin{equation}
y_j=y_j^{(n-1)}y_j^{(n-2)} \cdots y_j^{(0)}
\label{y_j}
\end{equation}

Therefore, the problem of constructing the quantum circuit for $\mathcal{U}_k$ turns into finding a logical function $f_k:x_k^{(n-1)} x_k^{(n-2)}\cdots x_k^{(0)}\rightarrow y_k^{(n-1)} y_k^{(n-2)} \cdots y_k^{(0)}$. We refer to the design method of reversible logic circuit that can map the permutation matrix into a series of controlled Pauli-X, Pauli-Z gates with multiple control qubits. 

\begin{figure}
	\centering
		\includegraphics[width=16cm]{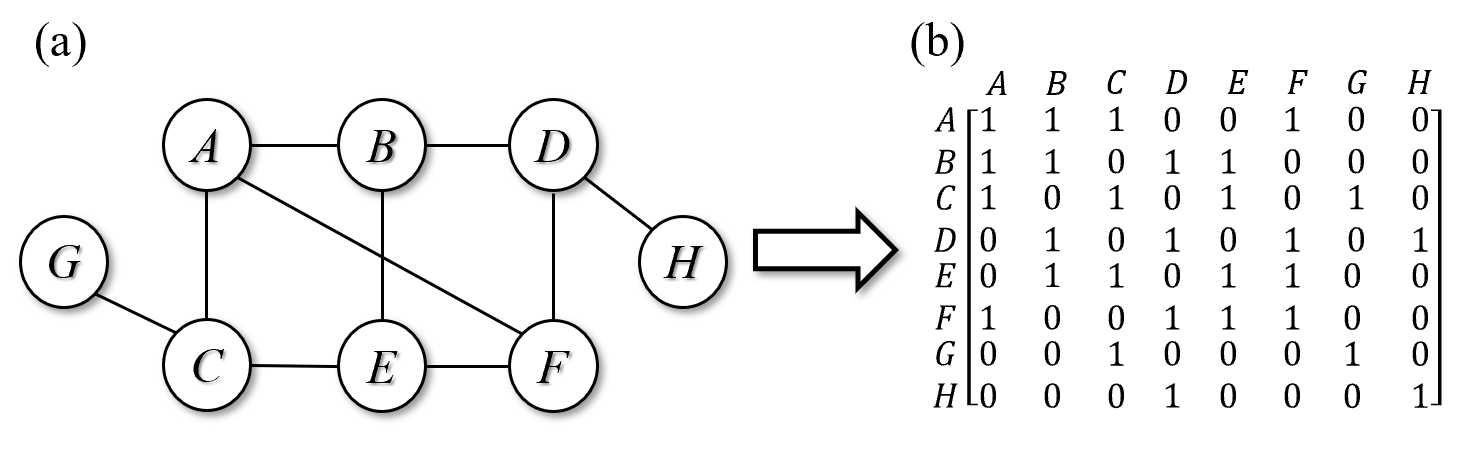}
	\caption{(a) An undirected graph with 8 nodes. (b) The corresponding adjacent matrix with self-connections.}
	\label{demo}
\end{figure}

By stacking all the sub-quantum circuits of $\mathcal{U}_k$, we can approximate the quantum circuit of the adjacent matrix $\hat{A}$. In order to visually demonstrate how to deploy an adjacent matrix onto a quantum circuit, we take a graph with 8 nodes as an example, as shown in \textcolor{red}{Fig \ref{demo}(a)}. To simplify the demonstration, we omit the normalization process and only discuss finding its corresponding adjacent matrix $\widetilde{A}$, as shown in \textcolor{red}{Fig \ref{demo}(b)}. 

\begin{figure}
	\centering
		\includegraphics[width=16cm]{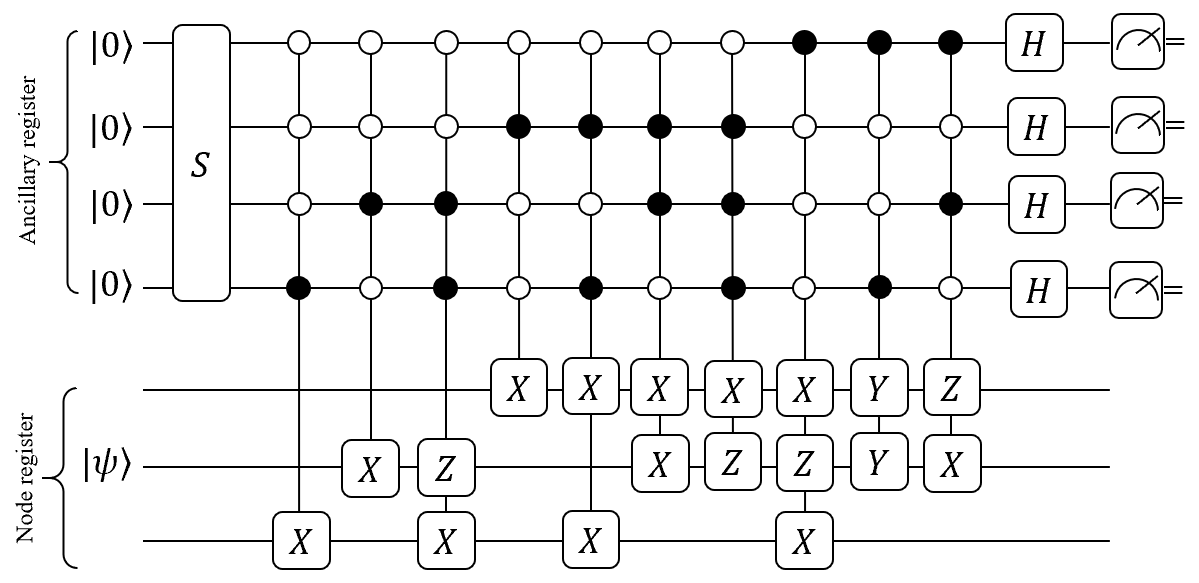}
	\caption{ The quantum circuit for performing the adjacent matrix $\widetilde{A}$, when the four ancillary qubits are observed are 0. $S$ is a unitary operator where the first column is $[1,1,-1,2,-1,1,1,1,1,1,1,0,0,0,0,0]/\sqrt{14}$, and other columns can be constructed by Gram-Schmidt orthogonalization~\textcolor{red}{\cite{Gram-Schmidt}}}
	\label{LCUdecomposed}
\end{figure}

According to the \textbf{Approach 1}, the adjacent matrix $\widetilde{A}$ can be constructed by 11 combinations of Pauli operators: $\{III$, $IIX$, $IXI$, $IXZ$, $XII$, $XIX$, $XXI$, $XZI$, $XZX$, $YYI$, $ZXI\}$ (omitted the symbol of Kronecker product '$\otimes$'). Based on LCU, we can use an ancillary register with 4 qubits and a node register with 3 qubits to perform the adjacent matrix $\widetilde{A}$, as shown in \textcolor{red}{Fig \ref{LCUdecomposed}}.

According to \textbf{Approach II}, the adjacent matrix $\widetilde{A}$ is decomposed to 4 permutation matrices, $\widetilde{A}=\sum_{k=0}^3 \mathcal{U}_k$, and each $\mathcal{U}_k$ is shown as follows:
\begin{equation}
\begin{split}\mathcal{U}_0=
\begin{bmatrix}
1 & 0 & 0 & 0 & 0 & 0 & 0 & 0 \\
0 & 1 & 0 & 0 & 0 & 0 & 0 & 0 \\
0 & 0 & 1 & 0 & 0 & 0 & 0 & 0 \\
0 & 0 & 0 & 1 & 0 & 0 & 0 & 0 \\
0 & 0 & 0 & 0 & 1 & 0 & 0 & 0 \\
0 & 0 & 0 & 0 & 0 & 1 & 0 & 0 \\
0 & 0 & 0 & 0 & 0 & 0 & 1 & 0 \\
0 & 0 & 0 & 0 & 0 & 0 & 0 & 1 
\end{bmatrix};
\mathcal{U}_1=
\begin{bmatrix}
0 & 1 & 0 & 0 & 0 & 0 & 0 & 0 \\
1 & 0 & 0 & 0 & 0 & 0 & 0 & 0 \\
0 & 0 & 0 & 0 & 0 & 0 & 1 & 0 \\
0 & 0 & 0 & 0 & 0 & 0 & 0 & 1 \\
0 & 0 & 0 & 0 & 0 & 1 & 0 & 0 \\
0 & 0 & 0 & 0 & 1 & 0 & 0 & 0 \\
0 & 0 & 1 & 0 & 0 & 0 & 0 & 0 \\
0 & 0 & 0 & 1 & 0 & 0 & 0 & 0 
\end{bmatrix}; \\
\mathcal{U}_2=
\begin{bmatrix}
0 & 0 & 1 & 0 & 0 & 0 & 0 & 0 \\
0 & 0 & 0 & 1 & 0 & 0 & 0 & 0 \\
0 & 0 & 0 & 0 & 1 & 0 & 0 & 0 \\
0 & 0 & 0 & 0 & 0 & 1 & 0 & 0 \\
0 & 1 & 0 & 0 & 0 & 0 & 0 & 0 \\
1 & 0 & 0 & 0 & 0 & 0 & 0 & 0 \\
0 & 0 & 0 & 0 & 0 & 0 & 1 & 0 \\
0 & 0 & 0 & 0 & 0 & 0 & 0 & 1 
\end{bmatrix};
\mathcal{U}_3=
\begin{bmatrix}
0 & 0 & 0 & 0 & 0 & 1 & 0 & 0 \\
0 & 0 & 0 & 0 & 1 & 0 & 0 & 0 \\
1 & 0 & 0 & 0 & 0 & 0 & 0 & 0 \\
0 & 1 & 0 & 0 & 0 & 0 & 0 & 0 \\
0 & 0 & 1 & 0 & 0 & 0 & 0 & 0 \\
0 & 0 & 0 & 1 & 0 & 0 & 0 & 0 \\
0 & 0 & 0 & 0 & 0 & 0 & -1 & 0 \\
0 & 0 & 0 & 0 & 0 & 0 & 0 & -1 
\end{bmatrix};
\end{split}
\end{equation}

\begin{figure}
	\centering
		\includegraphics[width=16cm]{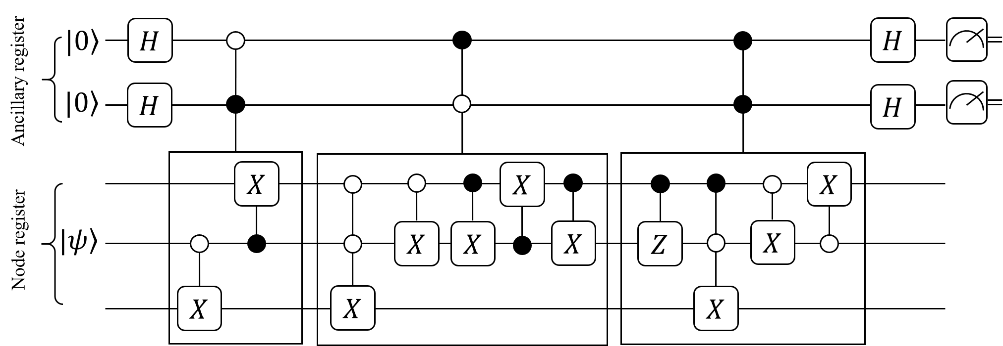}
	\caption{ The quantum circuit of the adjacent matrix $\widetilde{A}$, based on LCU and the design method of reversible logical circuit.}
	\label{QKMdecomposed}
\end{figure}

\noindent Based on LCU, we can use an ancillary register with 2 qubits and a node register with 3 qubits to perform the adjacent matrix $\widetilde{A}$, as shown in~\textcolor{red}{Fig \ref{QKMdecomposed}}.

Finally, we successfully obtain $\widetilde{A}\ket{\psi}$, when the measurement result of the ancillary register is 0.  

However, the unitary matrix factorization of a non-unitary matrix is  far less straightforward than in the aforementioned example. The decomposition method like Equation~\eqref{U_k} is not unique, thus the problem is not only how to obtain each specific $h_k\mathcal{U}_k$, but also how to find the optimal decomposition method. Then, we can use \textbf{Approach 1} and \textbf{Approach 2} to construct a quantum circuit accordingly. 

We regard optimally decompose the $\hat{A}$ for the quantum circuit constructing as a multi-objective optimization problem. The three optimization targets are as follows:
\begin{enumerate}
\item [(1)] to minimize the gate complexity of each $\mathcal{U}_k$;

\item [(2)] to maximum the successful probability of LCU $p$;

\item [(3)] to minimize the Frobenius norm of $\Delta$, denoted as $||\Delta||$.
\end{enumerate}	

\noindent For this multi-objective optimization problem, we take the second-generation of non-dominated sorting genetic algorithms (NSGA-II) to searching the mapping relationship between the given adjacent matrix and the quantum circuit \textcolor{red}{\cite{NSGA-II}}.

We achieve the searching procedure on a  classical computer. However, it is possible to move this part onto quantum circuit in the future, as quantum searching algorithm is mature. In the worst case, the complexity of the searching is $O(M2^n!)$, where $M$ is the number of $\mathcal{U}_k$, and $n$ is the number of the qubit. Once a mapping relationship between a given graph and a quantum circuit is determined, we can reuse the quantum circuit no matter how the features of the nodes change. In addition, 
as an approximation accuracy requirements decreases (i.e., a larger $\Delta$ is allowed), the computational complexity will be greater reduced.

\subsection{Adjustable Quantum Gates to Replace the Weight Matrix}
\begin{figure}
	\centering
		\includegraphics[width=12cm]{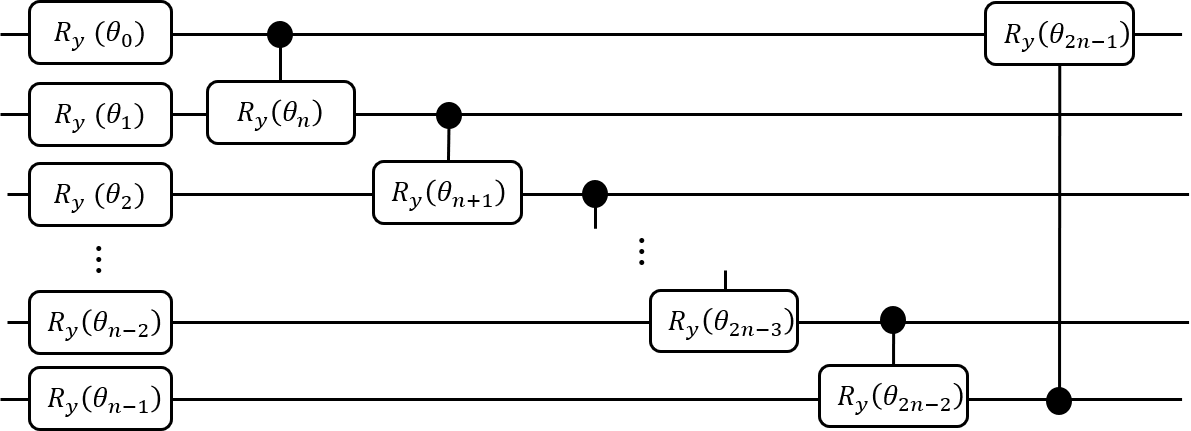}
	\caption{The architecture of PQC block.}
	\label{Block}
\end{figure}

We add an $Q$ operator (i.e. a set of adjustable quantum gates) after the quantum circuit for the adjacent. The role of $Q$ operator is to replace the tunable weight matrix in GCL. According to the previous studies on PQC~\textcolor{red}{\cite{zoufal2019quantum}}, we use a series of $R_y(\theta)$ gates to build a representative PQC block, as shown in \textcolor{red}{Fig \ref{Block}}. 

\subsection{Quantum Graph Convolutional Network}
\label{sec_QGCN}
\begin{figure}
	\centering
		\includegraphics[width=10cm]{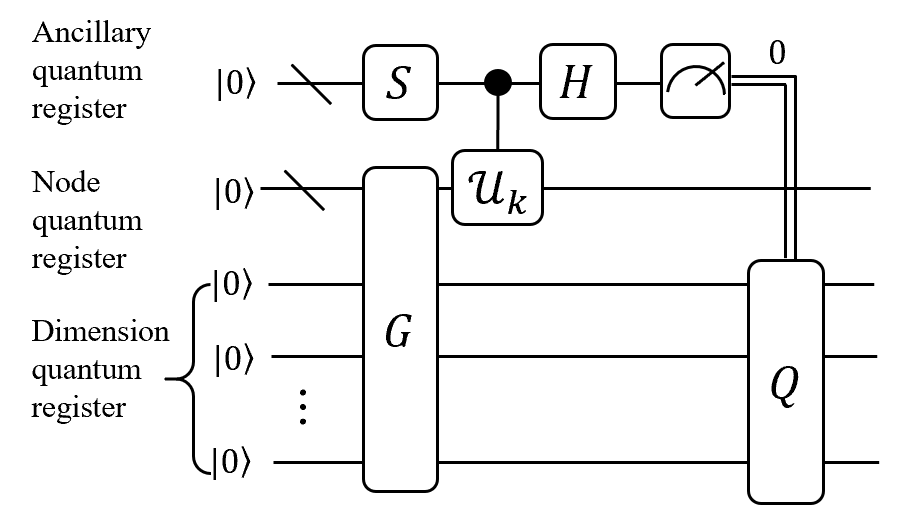}
	\caption{The quantum circuit of a QGCL, where $G$ is the operator for preparing the quantum initial state, $S$ is a unitary operator that the first column is $h_k$, and $Q$ is a set of adjustable quantum gates.}
	\label{qcQGCL}
\end{figure}

\begin{figure}
	\centering
		\includegraphics[width=16cm]{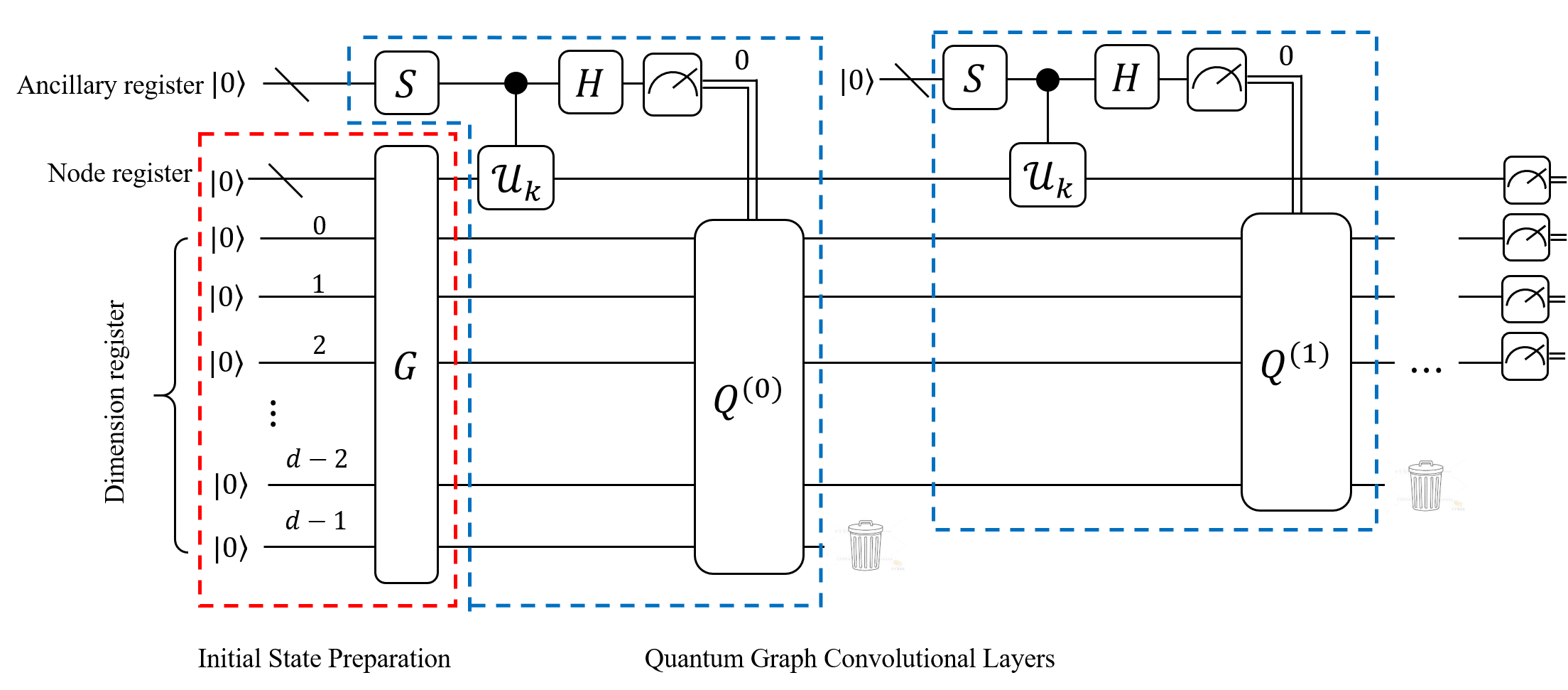}
	\caption{The quantum circuit of QGCN.}
	\label{QGCN}
\end{figure}

After constructing the adjacent matrix and the weight matrix, we construct the quantum circuit for QGCL, as shown in \textcolor{red}{Fig \ref{qcQGCL}}. The quantum state evolution is shown in Equation \eqref{evaQGCL}. Subsequently, we stack multiple QGCLs to obtain the QGCN model, as shown in \textcolor{red}{Fig \ref{QGCN}}. Note that since the dimension of input is usually higher than that of output, in QGCN, we discard some qubits of dimension register to guarantee a weight matrix square. 
 
\begin{equation}
\begin{split}
\ket{0} \otimes \ket{0} \otimes \ket{0} \xrightarrow{G} \ket{0} \otimes \sum_{j=0}^{N-1} \sum_{d=0}^{D-1} \mu_{j,d}\ket{j}\otimes\ket{d} \\
\xrightarrow{S} \left(\sum_{k=0}^{M-1}h_k\ket{k}\right) \otimes \left(\sum_{j=0}^{N-1} \sum_{d=0}^{D-1} \mu_{j,d}\ket{j}\otimes\ket{d} \right) \\
\xrightarrow{\mathrm{Controlled} \;\; \mathcal{U}_k} \sum_{k=0}^{M-1} h_k\ket{k} \otimes \left(\mathcal{U}_k \otimes I^{\log_2 D}\right) \left(\sum_{j=0}^{N-1} \sum_{d=0}^{D-1} \mu_{j,d}\ket{j}\otimes\ket{d} \right)\\
\xrightarrow[\mathrm{\;\;the\;\;ancillary\;\;register\;\;is\;\;0}]{H \;\; \mathrm{and\;\;the\;\;measurement\;\;of}} \left(\hat{A} \otimes I^{\log_2 D} \right) \left(\sum_{j=0}^{N-1} \sum_{d=0}^{D-1} \mu_{j,d}\ket{j}\otimes\ket{d}\right) \\
\xrightarrow{Q}\left(I^{\log_2 N} \otimes Q \right)\left(\hat{A} \otimes I^{\log_2 D} \right)\left(\sum_{j=0}^{N-1} \sum_{d=0}^{D-1} \mu_{j,d}\ket{j}\otimes\ket{d}   \right)
\end{split}
\label{evaQGCL}
\end{equation}

\subsection{Training Process of QGCN}
We utilize the gradient descent algorithm to update the learning parameters. The dimension of Hilbert space increases exponentially with the number of qubits, therefore, it is difficult to compute gradients of large-scale matrices on a classical computer. This is the main reason in the present work for us to use quantum circuits to compute the gradient of QGCN.

The PSR provides a method to compute the gradient of a parameterized quantum gate (PQG) $U_B(\theta)=e^{-i \theta B}$ \textcolor{red}{\cite{psr}}. A constraint of using PSR is that $B$ is a unitary and Hermitian matrix, i.e. $BB=I$. According to the PSR, after applying $U_B (\theta)$, the measurement expectation $E(\theta)$ is shown in Equation \eqref{psr_UG}.

\begin{equation}
E(\theta)=\braket{\varphi|U_B^{\dagger}(\theta)O U_B|\varphi}
\label{psr_UG}
\end{equation}

\noindent Then the gradient of $\theta$ is shown in Equation \eqref{partial_UG}.

\begin{equation}
\frac{\partial E(\theta)}{\partial \theta}=E\left(\theta+\frac{\pi}{4}\right)-E\left(\theta-\frac{\pi}{4}\right)
\label{partial_UG}
\end{equation}

\noindent Because a (controlled) $R_y (\theta)=e^{-i\theta Y}$ gate is a fundamental ceil to construct a PQC and a Pauli-$Y$ operator and it satisfies the constraint of PSR, we are able to use PSR to compute the gradient of QGCN. We demonstrate that PSR can be applied to the entire QGCN by illustrating how to use PSR on a QGCL.

As shown in \textcolor{red}{Fig \ref{QGCN}}, the node register stores the topological information and the dimension register stores the dimension information of all nodes. Under the premise that the measurement result of the ancillary register is 0 given the measurement result of the node register is $j$, we maximize the expectation function $E_{j,y_j} (\vec{\theta})$, where $E_{j} (\vec{\theta})$ is a column vector and its size equals the output dimension; $y_j$ denotes the label of the $j$-th node, is a real number. For multi-classification problems, we use cross entropy as the loss function $\mathcal{L}_c (\vec{\theta})$ and minimize the loss function in the train.

\begin{equation}
\mathcal{L}_c (\vec{\theta})=
-\frac{1}{N} \sum_{j=0}^{N-1} \sum_{\iota=0}^{D^{(1)}-1}
\mathrm{vec}_{\iota}(y_j) \cdot \ln\left(\mathrm{softmax}_\iota\left(E_j(\vec{\theta})\right)\right)
\end{equation}

\noindent where $D^{(1)}$ is the total number of categories in a dataset has a total of  categories. $\mathrm{vec}(y_j)$ is a one-hot column vector with $D^{(1)}$ dimensions, where the $y_j$-th element is 1; $\mathrm{vec}_\iota(y_j)$ denotes the $\iota$-th element of $\mathrm{vec}(y_j)$. $E_{j,\iota} (\vec{\theta})$ represents the predicted probability that the $j$-th node belongs to the $\iota$-th category.

We first focus on the derivative of  $E_{j,\iota} (\vec{\theta})$. For simplification, as $\left(H \otimes I^{ \otimes \log_2 N +\log_2 D}\right)$ $\left(\mathrm{C}-\mathcal{U}_k \otimes I^{\otimes \log_2 D}\right)$ $(S \otimes G)$ is fixed in a QGCN, we use $\mathcal{U}$ to replace it. 

\begin{equation}
E_{j,\iota}(\vec{\theta})=
\braket{0|\left(Q(\vec{\theta})\mathcal{U}\right)^\dagger O_{j,\iota}\left(Q(\vec{\theta})\mathcal{U}\right)|0}
\end{equation}

\noindent where, $O$ represents the observation operator. In mathematics, the observation operator can be regarded as a diagonal matrix with $2^{a+\log(\lceil N \rceil)+\log(\lceil D^{(1)} \rceil)}$ dimension, where $a$ is the number of ancillary qubits, $N$ is the number of nodes in the graph, and $D^{(1)}$ is the output dimension of each node. $\vec{\theta}$ is the vector of tunable parameters, which has $T$ elements, in total.  More specifically, $O_{j,\iota}$ represents that the $(j2^{\lceil D^{(1)} \rceil}+\iota)$-th element is 1 on the diagonal of the observation operator. $E_{j,\iota}(\vec{\theta})$ denotes the $\iota$-th element of the column vector $E_{j}(\vec{\theta})$. The gradient of $\theta_\tau$ is shown in Equation \eqref{partial_E}.

\begin{equation}
\begin{split}
\frac{\partial E_{j,\iota}(\vec{\theta})}{\partial \theta_\tau}= \\
\braket{0| \left(\prod_{t=0}^{\tau-1}Q(\theta_t)Q\left(\theta_\tau+\frac{\pi}{4}\right)\prod_{t=\tau+1}^{T-1}Q(\theta_t) \mathcal{U}\right)^\dagger O_{j,\iota} \left(\prod_{t=0}^{\tau-1}Q(\theta_t)Q\left(\theta_\tau+\frac{\pi}{4}\right)\prod_{t=\tau+1}^{T-1}Q(\theta_t) \mathcal{U}\right) |0}\\
-\braket{0| \left(\prod_{t=0}^{\tau-1}Q(\theta_t)Q\left(\theta_\tau-\frac{\pi}{4}\right)\prod_{t=\tau+1}^{T-1}Q(\theta_t) \mathcal{U}\right)^\dagger O_{j,\iota} \left(\prod_{t=0}^{\tau-1}Q(\theta_t)Q\left(\theta_\tau-\frac{\pi}{4}\right)\prod_{t=\tau+1}^{T-1}Q(\theta_t) \mathcal{U}\right) |0}
\end{split}
\label{partial_E}
\end{equation}

\noindent According to the compound function derivation chain rule, the derivation of loss function is shown in  Equation \eqref{partial_L}.

\begin{equation}
\frac{\partial \mathcal{L}_{c}(\vec{\theta})}{\partial \theta_\tau}= \frac{1}{N} \sum_{j=0}^{N-1}\sum_{\iota=0}^{D^{(1)}-1} \left(
\mathrm{softmax}_\iota \left(E_j(\vec{\theta})\right)-\mathrm{vec}_\iota(y_j)\right) \frac{\partial E_{j,\iota}(\vec{\theta})}{\partial \theta_\tau}
\label{partial_L}
\end{equation}
\noindent The tunable parameter $\theta_\tau$ can be updated according to Equation \eqref{update_theta}. 

\begin{equation}
\theta_\tau \leftarrow \theta_\tau - \frac{\partial \mathcal{L}_{c}(\vec{\theta})}{\partial \theta_\tau}
\label{update_theta}
\end{equation}

According to the aforementioned procedures, we successfully use gradient PQCs to calculate the gradient of each tunable parameter $\theta_\tau$, and thus achieve the gradient decent for a QGCL. 


\section{Numerical Simulation}
To verify the feasibility and robustness of a QGCN, we use the QGCN model to conduct a classification task based on Cora~\textcolor{red}{\cite{GCN}}, a real-world dataset with topological data. The Cora contains 2708 scientific publications and 5429 citation relationships, equivalent to 2708 nodes (140 nodes are for the training and others are for the test) and 5249 edges. These scientific publications are grouped into 7 categories based on article themes. Each publication in the dataset is represented by a feature vector. An element in a vector represents the absence/presence of the corresponding word in the dictionary. The dictionary consists of 1433 tokens, thus the length of a feature vector is 1433. For the Cora dataset, the number of input and output dimension is 1433 and 7, respectively. 

Currently, it is hard to simulate QGCN with two QGCLs. For the first layer, to use a QGCL, it requires 13 qubits and 11 qubits for the node quantum register (the Cora graph has 2708 nodes) and the dimension quantum register (the input dimension is 1433), respectively. The number of the qubits required challenges the computing power of a classical computer. To solve this problem, we replace the first layer using a classical QCL. In this way, as the data dimension is reduced to 16 after the first layer, only 4 qubits are additionally required to simulate a QGCL as a second layer, and the simulation becomes feasible.

In our simulation, we do two evaluations on the performance of QGCN:
\begin{itemize} 
\item [(1)] We evaluate whether QGCN has the same performance as GCN~\textcolor{red}{\cite{GCN}} and compare the required parameters under the same performance. 
\item [(2)] We evaluate the impact of $e_\Delta$ (i.e., the precision of the quantum adjacent matrix) on the accuracy of node classification.
\end{itemize}

\begin{table}[h] 
\renewcommand\arraystretch{1.5}
\setlength{\abovecaptionskip}{0.cm}
\caption{The simulation settings and the summary of the results.}
\begin{tabular}{c|ccc}
\hline
\hline

Model Name & \makecell*[c]{The number of tunable \\parameters in the \\ second QGCL} & \makecell*[c]{The error propor-\\tion of the residual \\ term $e_\Delta$} & \makecell*[c]{The node \\ classification \\ accuracy of the test set}\\
\hline
GCN & $16 \times 7$ & 0 & 0.792  \\
\hline
\multirow{5}{*}{\makecell*[c]{QGCN \\ with different \\ number of blocks}} & $8 \times 1$ & 0 & 0.499 \\
& $8 \times 2$ & 0 & 0.536 \\
& $8 \times 5$ & 0 & 0.733 \\
& $8 \times 10$ & 0 & 0.794 \\
& $8 \times 14$ & 0 & 0.792 \\
\hline
\multirow{5}{*}{\makecell*[c]{QGCN \\ with different \\ residual terms}} & $8 \times 10$ & 0.01 & 0.782 \\
& $8 \times 10$ & 0.05 & 0.752 \\
& $8 \times 10$ & 0.1 & 0.747 \\
& $8 \times 10$ & 0.2 & 0.716 \\
& $8 \times 10$ & without $\hat{A}$ & 0.692 \\
\hline
\hline
\end{tabular}
\label{settings}
\end{table}

\begin{figure}
  \centering
  \subfigure[]{
  \includegraphics[width=7.2cm]{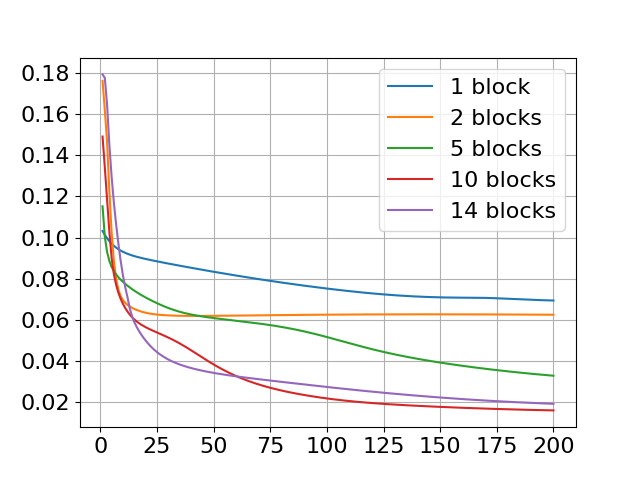}
  }
  \quad
  \subfigure[]{
  \includegraphics[width=7.2cm]{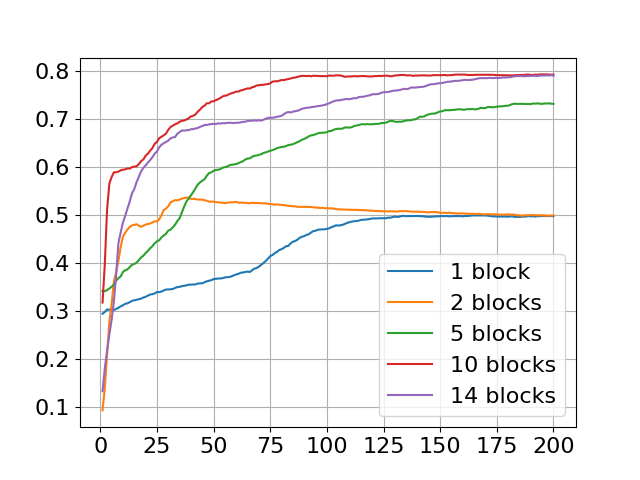}
  }
  \quad
  \subfigure[]{
  \includegraphics[width=7.2cm]{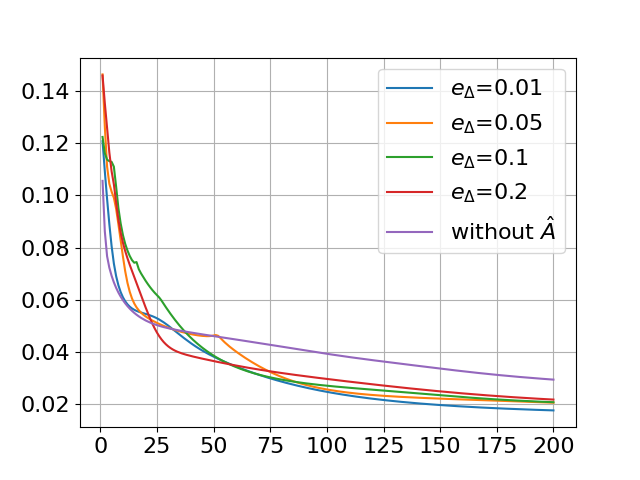}
  }
  \quad
  \subfigure[]{
  \includegraphics[width=7.2cm]{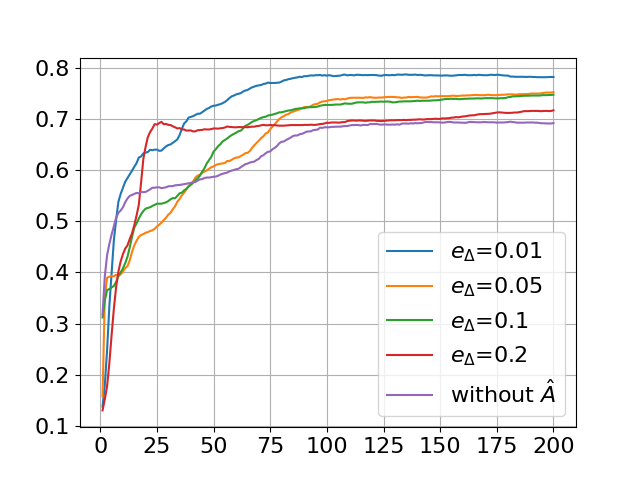}
  }
\caption{The simulation result of QGCN under the various size of parameters and various $\Delta$. The loss (a) and the classification accuracy (b) are shown as blue, yellow, green, red, and purple curves, corresponding to the blocks with the quantity of 1, 2, 5, 10, and 14, respectively. The loss values (c) and the classification accuracy (d) are shown as blue, yellow, green, red, and purple curves, corresponding to $e_\Delta$ of 0.01, 0.05, 0.1, 0.2, and without adjacent matrix $\hat{A}$, respectively.}
\label{result}
\end{figure}

Due to computing power limitation, in this numerical simulation, we use a classical GCL (see \textcolor{red}{Sec \ref{sec_GCL}}) as the first layer of QGCN (where the weight matrix in the first layer is  $W^{(0)} \in \mathbb{R}^{1433 \times 16}$, and QGCL (see \textcolor{red}{Sec \ref{sec_QGCN}}) as the second layer of QGCN. In the second layer, the number of the PQC blocks and the residual terms are set to be various. The specific simulation settings are listed in  \textcolor{red}{Table \ref{settings}}

\textbf{Evaluation 1.} In the second layer of QGCN, i.e., the QGCL, the size of the tunable phases is $8 \times 1$, $8 \times 2$, $8 \times 5$, $8 \times 10$, or $8 \times 14$ (a size denotes that tunable phases in a block times the number of the blocks).  The results, as shown in the upper part of \textcolor{red}{Table 1}, show that when the size is $8 \times 10$, QGCN achieves the same node classification accuracy as the classical GCN. However, the accuracy does not  further improve as the 
size of parameters grows. We can conclude that the QGCN achieves the same performance as the GCN yet with smaller size of parameters.

\textbf {Evaluation 2.} In a real experimental environment, the adjacent matrix can not be perfectly implemented on a quantum circuit, due to a high complexity of the adjacent matrix and  experimental noises. We can simulate the impact of the former reason, as it reflects on a $\Delta$. We define an error proportion $e_\Delta$ to measure the inaccuracy of the approximation to the adjacent matrix. For example, $e_\Delta=0.01$ denotes that after normalization, each element to represent each edge has 0.01 of the probability to be randomly added or subtracted 0.01, in \textcolor{red}{Table 1}. In the worst case, we remove the adjacent matrix on the quantum circuit and only use the PQC-based weight matrix.

The results, as shown in the lower part of \textcolor{red}{Table 1}, show that the accuracy decrease as $e_\Delta$ grows. When we remove the adjacent matrix on the quantum circuit, the accuracy becomes the minimum. Therefore, we can conclude that compared with the traditional PQC architectures, our QGCN efficiently improve the classification accuracy of the topological data.

\section{Discussion}
\begin{figure}
	\centering
		\includegraphics[width=10cm]{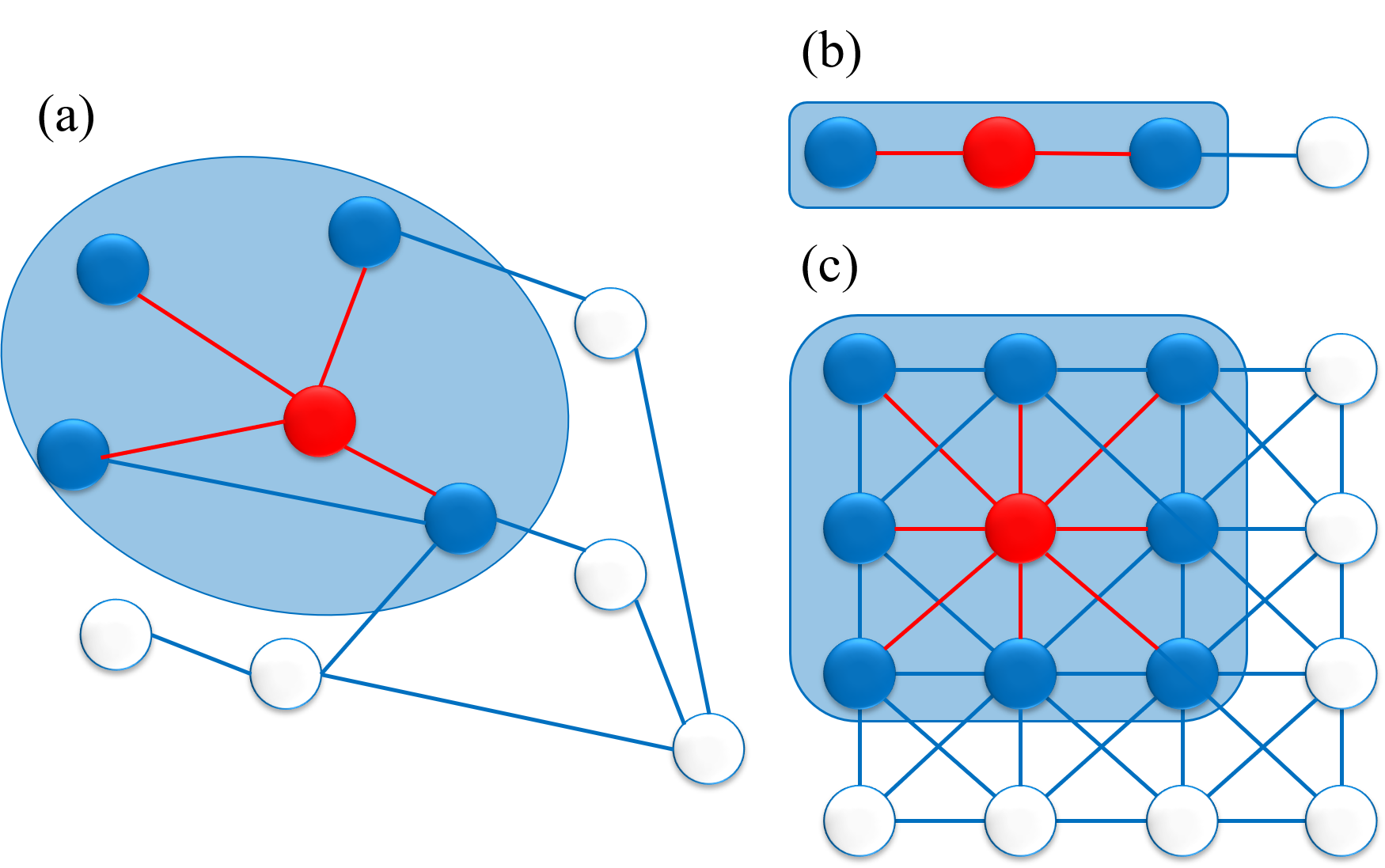}
	\caption{The topology graphs of (a)citation network, (b)1-D audio signal, (c)the digital image.}
	\label{1D_2D}
\end{figure}

We generalize the application range of QGCL, and we demonstrate that the novel PQC structure of QGCL can be used not only to realize the function of GCN, but also to realize the function of convolutional neural network (CNN). This is because both CNN and GCN can be regarded as aggregating the node features according to the topology. The advantage to use QGCN to do a CNN-based task is that it can adapt multi-dimensional tensors efficiently and use less tunable parameters. For example, audio signals and digital images can be viewed as 1D and 2D topological graphs, respectively. A frame or a pixel can be regarded as a node. Thus, each node of the 1D graph has 2 neighbor nodes (except the start and the end nodes), as shown in  \textcolor{red}{Fig \ref{1D_2D}(b)};  each node of the 2D graph has 2 neighbor nodes (except the surrounding and corner nodes), as shown in \textcolor{red}{Fig \ref{1D_2D}(c)}. We write the adjacent matrix of an audio signal with $N$ frames (Equation \eqref{adj_1D}) and a digital image with $N \times N$ pixels (Equation \eqref{adj_2D}).

\begin{equation}
	\widetilde{A}_{1D}=\left[                 
	\begin{array}{ccccc}  
	1 & 1 & & &         \\
	1 & 1 & 1  &  &   \\
	& \ddots & \ddots & \ddots & \\
	&  &  1 &1 & 1\\
	& &  & 1 &  1     \\
	\end{array}
	\right]_{N\times N}
\label{adj_1D}
\end{equation}

\begin{equation}
\widetilde{A}_{2D}=\left[\begin{matrix}
 I &  & &   &  &      \\
 V_1 & V_2 & V_3  &      & &   \\
 & \ddots & \ddots & \ddots  & &\\
 &  & \ddots & \ddots &\ddots  &\\
 & & & V_1 & V_2 & V_3\\
   & &  & &  & I    \\
\end{matrix}\right]_{N^2\times N^2}
\label{adj_2D}
\end{equation}
Where,$I \in \mathbb{R}^{N \times N}$ is an identity matrix. $V_1,V_2,V_3$ is shown in Equation \eqref{V}.
\begin{equation}
	V_1=\left(                 
	\begin{array}{ccccc}  
	0 &  & & &         \\
	1 & 1 & 1  &  &   \\
	& \ddots & \ddots & \ddots & \\
	&  &  1 &1 & 1\\
	& &  & &  0     \\
	\end{array}
	\right)_{N\times N}
	V_2=\left(                 
	\begin{array}{ccccc}  
		1 &  & &   &     \\
		1 & 1 & 1  &  &  \\
		& \ddots & \ddots & \ddots& \\

		&  & 1 &1 & 1\\
		& & &  &  1     \\
	\end{array}
	\right)_{N\times N}  
	V_3=\left(
	\begin{array}{ccccc}  
	0 &  & &   &       \\
	1 & 1 & 1  &    &     \\
	& \ddots & \ddots & \ddots  & \\
	& &   1 &1 & 1\\
	& &   &  & 0     \\
	\end{array}
	\right)_{N\times N}.
\label{V}
\end{equation}

For an audio signal, the input matrix is $\mathcal{X} \in \mathbb{R}^{N \times D}$, where $N$ is the number of frames and $D$ is the number of channels.  The order of each column follows the time order of the audio signal. For a digital image, the input matrix is $\mathcal{X} \in \mathbb{R}^{N^2 \times D}$, where $N$ is the number of pixels in a column and $D$ is the number of channels(e.g., RGB). The pixels in the image are connected end to end in columns. Consequently, we can also state that our QGCN is the generalized model of the existing quantum convolutional neural network (QCNN)~\textcolor{red}{\cite{wei2022quantum}}. When acting as a QCNN, compared with~\textcolor{red}{\cite{wei2022quantum}}, our model supports to use  quantum circuits for gradient descent and the multi-channel inputs.

\section{Conclusion}
In this paper, we redesign the current PQC for a single GCL by stacking the quantum circuits of adjacent matrix and weight matrix. To implement the non-unitary adjacent matrix, we propose an optimization algorithm for finding the best unitary decomposition of the adjacent matrix, i.e., to decompose it into a summation of several unitary matrices and a residual term. Then, we apply the approximate adjacent matrix onto the quantum device based on LCU. Additionally, we adopt PSR to realize the gradients decent on the quantum device for a GCL. By stacking multiple GCLs, we implement the entire QGCN. We evaluate the performance of QGCN by simulation. The numerical results show that the QGCN has the same classification accuracy as the classical GCN, however requires less tunable parameters. QGCN outperforms the traditional PQC structure (without the adjacent matrix). We also evaluate the impact of $\Delta$ on the QGCN performance. The results suggested that a smaller $\Delta$ (i.e., a better approximation) gains a better performance. Therefore, we can draw a conclusion that QGCN based on a novel PQC architecture, is able to efficiently handle the classification task for topological quantum data. We expect our work to inspire any other novel design of PQC architectures, which borrows ideas from the classical neural architectures.

\section {Code Availability}
Code and datasets for the QSNN model can be obtained from a GitHub repository (https://github.com/yanhuchen/Quantum-Graph-Convolutional-Network).

\section{Reference}

\bibliographystyle{unsrt}

\bibliography{reference}

\end{document}